\documentclass[longauth, letter]{aa}

\usepackage{graphicx}
\usepackage{txfonts}
\usepackage{lscape}
\usepackage{color}

\begin{document}

\title{The CARMENES search for exoplanets around M dwarfs}
\subtitle{HD\,147379\,b: A nearby Neptune in the temperate zone
of an early-M dwarf}

    \author{A.\,Reiners\inst{1}      
     \and I.\,Ribas\inst{2}          
     \and M.\,Zechmeister\inst{1}    
     \and J.A.\,Caballero\inst{3,4}  
     \and T.\,Trifonov\inst{5}
     \and S.\,Dreizler\inst{1}
     \and J.C.\,Morales\inst{2}
     \and L.\,Tal-Or\inst{1}
     \and M.\,Lafarga\inst{2}
     \and A.\,Quirrenbach\inst{4}
% ...
}
  \institute{ Institut f\"ur Astrophysik, Georg-August-Universit\"at,
              Friedrich-Hund-Platz 1, D-37077 G\"ottingen, Germany\\
              \email{Ansgar.Reiners@phys.uni-goettingen.de}
         \and Institut de Ci\`encies de l'Espai (CSIC-IEEC), Campus UAB, c/ de Can Magrans s/n,
              E-08193 Bellaterra, Barcelona, Spain
         \and Centro de Astrobiolog\'ia (CSIC-INTA), Instituto Nacional de T\'ecnica Aeroespacial,
              Ctra. de Torrej\'on a Ajalvir, km 4, E-28850 Torrej\'on de Ardoz, Madrid, Spain
         \and Zentrum f\"ur Astronomie der Universt\"at Heidelberg, Landessternwarte,
              K\"onigstuhl 12, D-69117 Heidelberg, Germany
	 \and Max-Planck-Institut f\"ur Astronomie,
	      K\"onigstuhl 17, D-69117 Heidelberg, Germany
% ...
}


   \date{\today}


   \abstract{We report on the first star discovered to host a planet
     detected by radial velocity (RV) observations obtained within the
     CARMENES survey for exoplanets around M dwarfs. HD\,147379 ($V =
     8.9$\,mag, $M = 0.58 \pm 0.08$\,M$_{\odot}$), a bright M0.0\,V
     star at a distance of 10.7\,pc, is found to undergo periodic RV
     variations with a semi-amplitude of $K = 5.1\pm0.4$\,m\,s$^{-1}$
     and a period of $P = 86.54\pm0.06$\,d.  The RV signal is found in
     our CARMENES data, which were taken between 2016 and 2017, and is
     supported by HIRES/Keck observations that were obtained since
     2000. The RV variations are interpreted as resulting from a
     planet of minimum mass $m_{\rm p}\sin{i} = 25 \pm
     2$\,M$_{\oplus}$, 1.5 times the mass of Neptune, with an orbital
     semi-major axis $a = 0.32$\,au and low eccentricity ($e < 0.13$).
     HD\,147379\,b is orbiting inside the temperate zone around the
     star, where water could exist in liquid form. The RV time-series
     and various spectroscopic indicators show additional hints of
     variations at an approximate period of 21.1\,d (and its first
     harmonic), which we attribute to the rotation period of the
     star.}


   \titlerunning{HD\,147379\,b: A nearby Neptune in the temperate
zone of an early-M dwarf     }

   \maketitle
%

\section{Introduction}
\label{sect:Introduction}

Low-mass M-type stars have attracted increasing attention in the
exoplanet community over the past decade. The low masses and small
radii of M dwarfs make the detection of rocky planets less challenging
than for Sun-like stars, and the typical detection limit for radial
velocity (RV) surveys is on the order of 1\,m\,s$^{-1}$ , which
permits the
discovery of rocky planets in their habitable zones
\citep[e.g.,][]{2005AN....326.1015M, 2007AsBio...7...85S,
  2007AsBio...7...30T, 2013A&A...549A.109B,
  2016Natur.536..437A}. These stars are also of particular interest
because they are by far the most numerous \citep{2006AJ....132.2360H,
  2016AAS...22714201H}. In addition, the transit signal of a rocky
planet around such stars is within reach of ground-based telescopes of
small aperture \citep[e.g.,][]{2016Natur.533..221G,
  2017Natur.544..333D}, and they are being targeted by upcoming
photometry space missions such as \emph{TESS} and \emph{PLATO}.

Until now, RV measurements of M dwarfs have produced 81 planet
discoveries in total, many of them in multiple
systems.\footnote{http://exoplanet.eu} Only 20 of them are more
massive than 0.1\,M$_{\rm Jupiter}$, suggesting that M dwarfs host
fewer giant planets than solar-mass stars \citep{2008PASP..120..531C,
  2010PASP..122..905J}. While this may be a consequence of the hot
Jupiters being more frequent around hotter stars because more building
material is available \citep{2012A&A...541A..97M}, the trend has not
been confirmed so far in transit surveys
\citep[e.g.,][]{2012AJ....143..111J, 2016A&A...587A..49O}. Low-mass
stars are also suspected to favor multi-planet systems, which results
in an average of more than two planets per host star
\citep{2015ApJ...807...45D}. An exoplanet survey targeted on nearby M
dwarf stars therefore promises to detect many low-mass planets around
nearby stars for which the perspectives for a detailed investigation
and characterization are good.

M dwarf stars as targets for exoplanet searches have their specific
challenges. While the spectral coverage of typical visual RV
instruments is well suited for FGK-type stars, the much redder
spectral energy distribution of M-type stars requires red-optical and
near-infrared coverage for better efficiency. Additionally, M dwarfs
are typically active, and a wide (simultaneous) wavelength coverage is
therefore extremely valuable to distinguish between
wavelength-dependent activity signals and wavelength-independent
planetary signals in RV measurements. CARMENES
\citep{2014SPIE.9147E..1FQ} addresses these issues. We have been
conducting a dedicated survey of about 300 well-characterized M dwarfs
\citep{2017A&A...604A..87G, 2017arXiv171106576R, pre012} since January
2016 within Guaranteed Time Observations. The performance of the
instrument has been demonstrated and compared to others in a paper on
M-type stars known to host planets \citep{2017arXiv171001595T}. The
present paper is dedicated to the first star discovered by CARMENES to
host a planet.

In the following, we introduce the host star HD\,147379 with its basic
properties in Sect.\,\ref{sect:star}, describe our data from CARMENES and
HIRES/Keck in Sect.\,\ref{sect:data}, and present our results from the
analysis of the radial velocity measurements in combination with various
activity indicators in Sect.\,\ref{sect:results}. Our results are then
summarized in Sect.\,\ref{sect:conclusions}.

\section{HD\,147379}
\label{sect:star}

The star HD\,147379 (GJ\,617\,A, HIP\,79755, J16167+672S) is bright
($V=8.9$\,mag; $J=5.8$\,mag) and classified as M0.0\,V
\citep{2015A&A...577A.128A}; it is located at a distance of $d =
10.735\pm0.026$\,pc \citep{2016A&A...595A...2G}. This star forms a
common proper motion pair with a fainter companion (EW Dra, M3.0\,V,
$V=10.6$\,mag) at a projected separation of 1.07\,arcmin, or about
690\,au at the distance of the system
\citep{2007AJ....133..889L}. From the CARMENES data of HD\,147379, we
detect marginal Doppler broadening caused by rotation with
$\varv\,\sin{i} = 2.7\pm1.5$\,km\,s$^{-1}$
\citep{2017arXiv171106576R}. At this low value of $\varv\sin{i}$, we
cannot entirely exclude that not rotation, but other effects such as a
spectral mismatch between HD~147379 and the reference star caused the
additional broadening, which means that the value of $\varv\sin{i}$ is
essentially an upper limit.

The star shows mild chromospheric Ca\,H \& K emission with a median
$S$-index of 1.53 measured from the HIRES data
\citep{2017AJ....153..208B}. Butler and collaborators reported an
absence of H$\alpha$ emission in most of their spectra, but some
H$\alpha$ emission detections in 12 of their 30 spectra. We cannot
confirm this detection in our CARMENES data; all of our spectra show
H$\alpha$ in absorption. We note that \citet{2017ApJ...834...85N} also
listed H$\alpha$ emission for HD\,147379, but their reference for this
value in fact reported H$\alpha$ in absorption
\citep{2002AJ....123.3356G}.

Using the ROSAT all-sky survey X-ray flux of GJ~617\,A and B
\citep{1999A&A...349..389V} and the flux ratio derived from a later ROSAT HRI
pointing \citep{2004A&A...417..651S}, we compute an X-ray luminosity of
$L_{\rm X} = 10^{27.6}$\,erg\,s$^{-1}$ for HD~147379 (GJ~617\,A). From the
relation between X-ray activity and rotational period given in Eq.\,11 in
\citet{2014ApJ...794..144R}, we estimate the rotational period to be $P \approx
31$\,d. The uncertainty of this estimate is approximately $\pm 20$ days
because X-ray values of individual stars show a large scatter around the
rotation-activity relation.

For the stellar properties, we adopted the values in the top part of
Table\,\ref{table:1}.
The atmospheric parameters $T_{\rm eff}$ and [Fe/H] were determined by
fitting PHOENIX-ACES synthetic spectra \citep{2013A&A...553A...6H} to the
CARMENES spectra, as described in \citet{2016A&A...587A..19P}.  We collected
broad-band photometry from several surveys covering all parts of the spectral
energy distribution (SED) of the target
\citep{2016csss.confE.148C}. Integrating this photometric SED allowed us to
determine the luminosity $L$ as described in \citet{msccifuentes}. The radius
$R$ was calculated from our measured $T_{\rm eff}$ and $L$, while the mass was
obtained from the linear $M-R$ relation measured by \citet{msccasal}. From the
stellar radius and projected equatorial velocity, we can estimate the
rotational period of HD\,147379 to be $P_{\rm rot} / \sin{i} =
11^{+15}_{-5}$\,d ($f \sin{i} = 0.09\pm0.05$\,d$^{-1}$). If the star is seen
under a low inclination angle, $i$, the real value of $P_{\rm rot}$ will be
lower (faster rotation). As explained above, however, the detection of
spectral Doppler broadening is only marginal, which means that rotational
periods longer than 10--20\,d cannot be excluded. A rotation period of 10\,d
or longer is consistent with the absence of H$\alpha$ emission; faster
rotators ($P_{\rm rot} \le 10$\,d) typically show emission, while slower
rotators tend to lose their ability to generate it \citep{pre012}.

\section{Data}
\label{sect:data}

\begin{figure}
  \centering
  \includegraphics[width=\hsize]{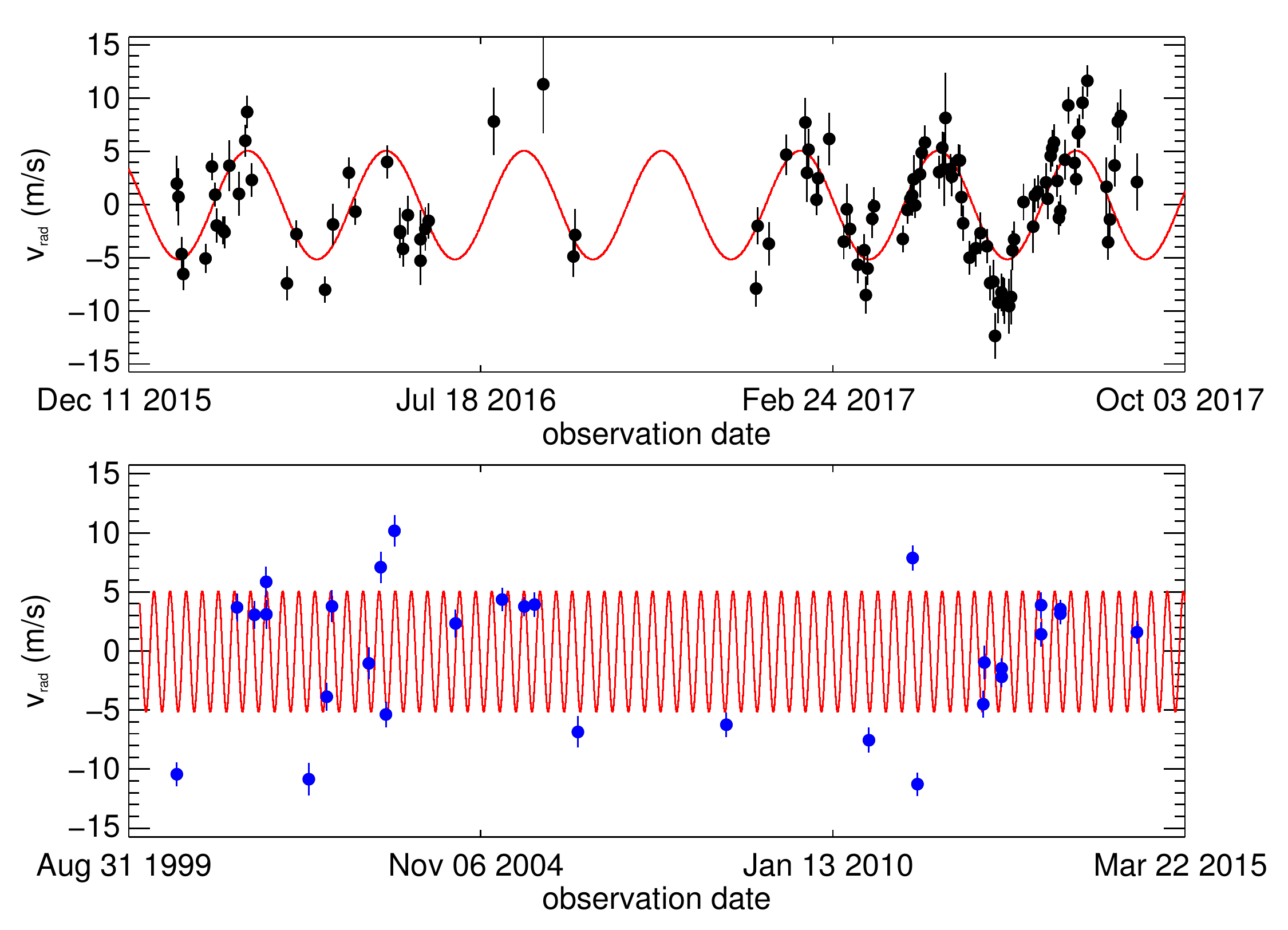}
  \caption{Radial velocities obtained with CARMENES (upper panel) and
    HIRES/Keck (lower panel). The orbital motion according to the adopted
    solution is overplotted in red (see Sect.\,\ref{sect:results}).}
  \label{fig:RVs}
\end{figure}

We analyzed data from the CARMENES VIS channel and HIRES/Keck. The
CARMENES measurements were taken in the context of the CARMENES search
for exoplanets around M dwarfs. The CARMENES instrument consists of
two channels: the VIS channel obtains spectra at a resolution of $R =
94,600$ in the wavelength range 520--960\,nm, while the NIR channel
yields spectra of $R = 80,400$ covering 960--1710\,nm. Both channels
are calibrated in wavelength with hollow-cathode lamps and use
temperature- and pressure-stabilized Fabry-P\'erot etalons to
interpolate the wavelength solution and simultaneously monitor the
spectrograph drift during nightly operations
\citep{2015A&A...581A.117B}.

Observations with CARMENES were tailored to obtain a signal-to-noise
ratio of 150 in the $J$ band, and the typical exposure time for HD~147379
was 7\,min.
The median internal RV precision of the CARMENES VIS channel exposures
of HD\,147379 is $\sigma_{\rm VIS} = 1.7$\,m\,s$^{-1}$. The
corresponding values of the internal RV precision in the NIR channel
are significantly higher, $\sigma_{\rm NIR} = 8.6$\,m\,s$^{-1}$,
mainly because in early-M dwarfs the amount of spectral features is
higher in the VIS channel spectral range
\citep{2017arXiv171106576R}. For the analysis carried out in this
paper, we therefore only used RVs from the VIS channel. From the
CARMENES data, the reduction pipeline provides information about
chromospheric emission from H$\alpha$, the variation of the line
profile shape (dLw), and the chromatic index (crx), as detailed in
\citet{pre018}.

We also computed the cross-correlation function (CCF) of each spectrum using a
weighted binary mask that was built from coadded observations of the star
itself. We selected $\sim$3000 deep, narrow, and unblended lines
that were
weighted according to their contrast and inverse full-width at half-maximum
(FWHM). We computed one CCF for each spectral order and subsequently combined
these individual CCFs according to the signal-to-noise ratio to compute the final
CCF. We fitted a Gaussian function to the central part of the combined CCF and
determined the radial velocity, FWHM, contrast, and bisector span. The latter
is defined as the difference between the average bisector values in the CCF
regions from 90\,\% to 60\,\% and from 40\,\% to 10\,\%.

We further included in our analysis the RV data from HIRES/Keck
published in \citet{2017AJ....153..208B}. Thirty observations are
reported between May 2000 and July 2014, and they have a median
internal uncertainty of $\sigma_{\rm Keck} = 1.1$\,m\,s$^{-1}$. We
show all data in Fig.\,\ref{fig:RVs}. \citet{2017AJ....153..208B}
reported a signal requiring confirmation at $P = 2.1$\,d.

\section{Results}
\label{sect:results}


\begin{figure}
  \centering
  \includegraphics[width=\hsize, bb = 30 0 425 645, clip=]{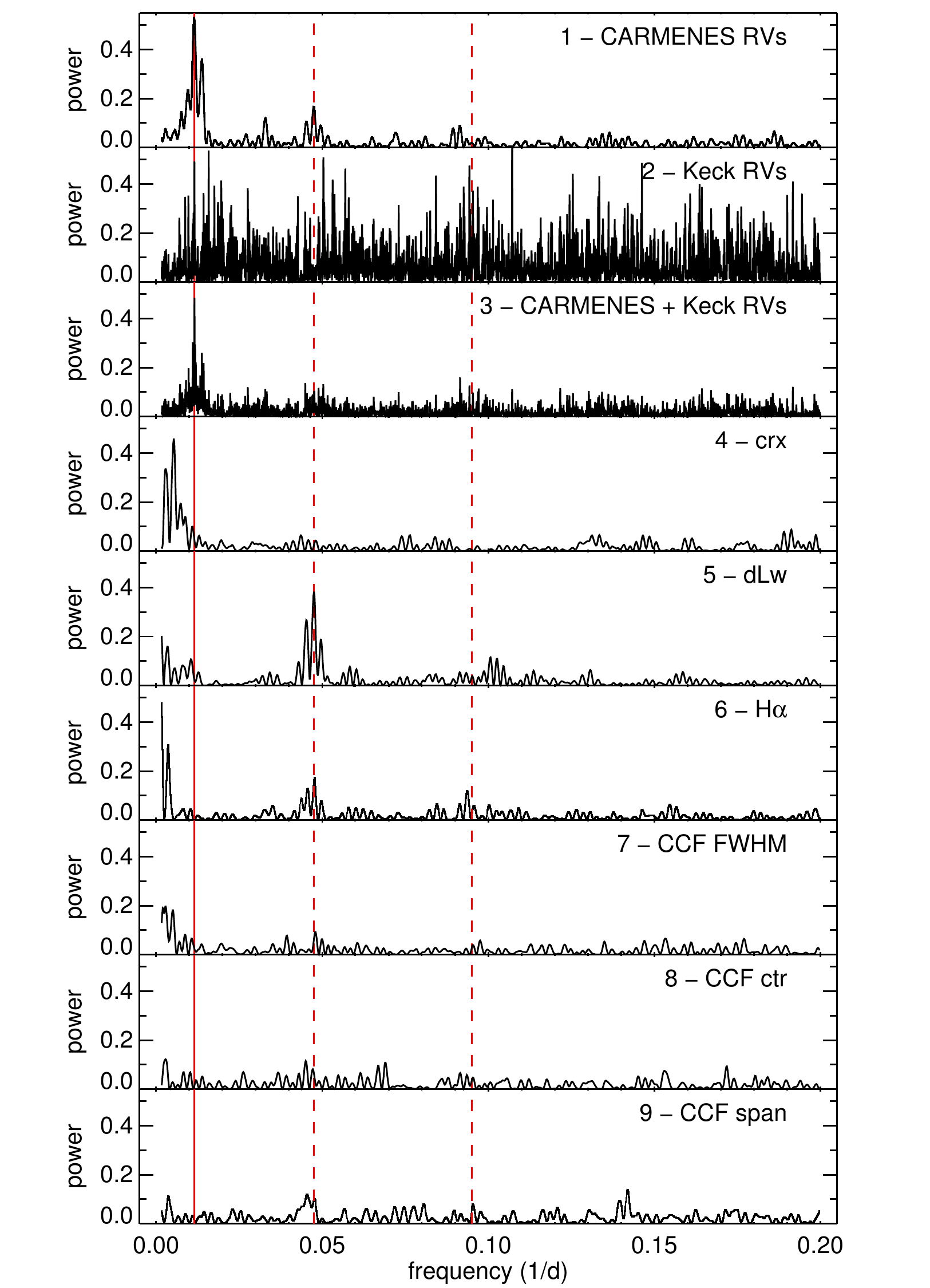}
  \caption{Periodograms from CARMENES data (top panel), HIRES/Keck data
    (second panel), and the combined data set (panel 3). The period of
    86.5\,d is marked with a vertical red line. Panels 4--9 from top
    to bottom show the chromatic index (crx), the differential line width
    (dLw), and the H$\alpha$ index as defined in \citet{pre018}, and the FWHM,
    the bisector contrast, and the bisector span from the CCF. Excess power at
    around $f=0.047$\,d$^{-1}$ in the CARMENES RVs and dLw is likely
    caused by stellar rotation (left red dashed line; the right dashed
    line shows its first harmonic at $f=0.094$\,d$^{-1}$).}
  \label{fig:Periodogram}
\end{figure}

We show periodograms using the generalized Lomb-Scargle formalism
\citep[GLS,][]{2009A&A...496..577Z} in Fig.\,\ref{fig:Periodogram}. In the top
panel, CARMENES RVs show a prominent signal at a period of $P = 86.5$\,d ($f =
0.0115$\,d$^{-1}$). Because of the limited total time baseline of CARMENES
observations, the peak is relatively broad. Keck observations cover a longer
time span, which leads to a higher frequency resolution in the periodogram, but
the number of observations is not sufficient to identify any statistically
significant peak. Nevertheless, we observe excess power at around $f =
0.0115$\,d$^{-1}$, which means that HIRES/Keck data are consistent with a
86.5 d periodicity. The periodogram from both data sets together reveals a
clear and unique signal at this frequency, as shown in the third panel of
Fig.\,\ref{fig:Periodogram}. To test whether the signal is persistent in the
CARMENES data, we calculated periodograms from the first and second half of the
CARMENES RV data alone. We found the peak at $f = 0.0115$\,d$^{-1}$ in both
cases. At $P = 2.1$\,d, the period where \citet{2017AJ....153..208B} reported
a signal from their data alone, the CARMENES and combined data sets do not
show any signal.

The CARMENES RV periodogram is relatively free of other significant peaks at
frequencies longer than $f = 0.02$\,d$^{-1}$ ($P < 50$\,d). An interesting
group of periodogram peaks appears around $f = 0.047$\,d$^{-1}$ ($P =
21.3$\,d). This feature may be connected to the rotational period of the
star. We investigated line profile indicators and the H$\alpha$ index as described
in \citet{pre018}. We note that for the H$\alpha$ index, we also calculated
values when we saw H$\alpha$ in absorption. Periodograms of the chromatic
index (crx), differential line width (dLw), and H$\alpha$ index are provided
in panels 4--6 in Fig.\,\ref{fig:Periodogram}, while panels 7--9 show
periodograms for the FWHM, the bisector contrast, and the bisector span from the CCF. None
of the six indicators show evidence for periodicity at $P = 86.5$\,d, but dLw
shows signicificant excess power at $P = 21.1$\,d. The H$\alpha$ index and
some of the three CCF indices show periodogram peaks around $P = 21.1$\,d, $P
= 10.6$\,d, or both, but no power at $P = 86.5$\,d.

From all the gathered evidence, we suspect that the secondary peak at $f =
0.047$\,d$^{-1}$ in the CARMENES RVs is caused by rotational modulation of the
star that is rotating at $P = 21.1$\,d (or $P = 10.6$\,d). We note that
21.1\,d and 10.6\,d are close to the fourth and eighth harmonics of
86.5\,d. However, if rotational modulation were the origin of these
periodogram peaks, we would also expect the first harmonic around 42\,d (and
others) to be prominent. Because we do not find this peak, and because we do
not find excess power in the activity indicators at $P = 86.5$\,d, we conclude
that this latter period is of planetary origin.

\begin{figure}
  \centering
  \includegraphics[width=\hsize]{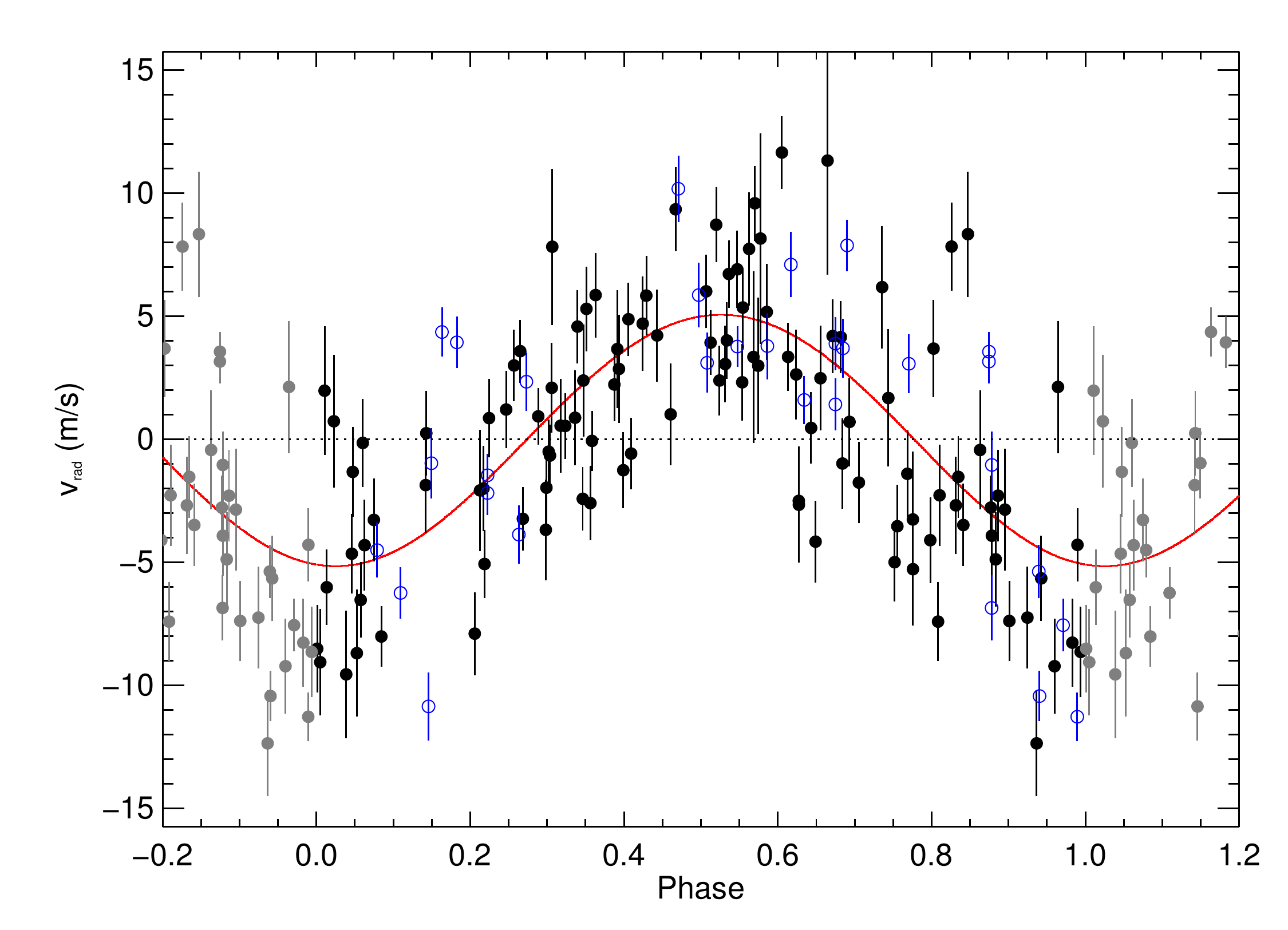}
  \caption{Phased RVs (black filled circles: CARMENES;
    blue open circles: HIRES/Keck). The red line shows the orbital
    motion caused by the planetary companion according to the solution
    in Table\,\ref{table:1}. Gray points are repetitions.}
  \label{fig:Phase_RV}
\end{figure}

We determined the parameters of the planet around HD\,147379 using a Keplerian
model coupled with a Nelder-Mead simplex algorithm \citep{NelderMead,Press},
which minimizes the negative logarithm of the model's likelihood function.  In
our modeling scheme, we also fit for the unknown RV jitter variance
$\sigma_{\rm jitter}$ of the HIRES/Keck and CARMENES data sets, which we
incorporated following the recipe in \citet{Baluev2009}. We quantified the
significance of the signal using the $\Delta \ln{L}$ statistic, as discussed in
\citet{2016Natur.536..437A}. When we used CARMENES data alone, we obtained $\Delta
\ln{L} = 41.7$, which results in a false-alarm probability (FAP) of $\sim
6.2\times 10^{-15}$. The inclusion of the HIRES/Keck data increased this value
to $\Delta \ln{L} = 51.2$ (FAP of $\sim 1.05\times 10^{-17}$), meaning that
the significance of the detection is further improved by these data. A search
for a second signal using the same method did not show evidence for any
periodic variability left in the data.

We estimated the uncertainties of the derived orbital parameters by
running the Markov chain Monte Carlo (MCMC) sampler {\tt emcee}
\citep{Mackey2013} in conjunction with our model, and as the
uncertainty,
we adopted the 68.3\% (1$\sigma$) credibility interval of the resulting
posterior parameter distribution. The results from our Keplerian
modeling of the HD\,147379 data are summarized in Table\,\ref{table:1},
together with the stellar parameters.

\begin{table}
  \caption{Parameters of HD\,147379 with 1$\sigma$ uncertainties}
  \label{table:1}
  \centering
  \begin{tabular}{ll}
    \hline\hline
    \noalign{\smallskip}
    Parameter & HD~147479\\
    \hline
    \noalign{\smallskip}
    $M$ (M$_{\odot}$) & $0.58 \pm 0.08$\\    
    $R$ (R$_{\odot}$) & $0.57 \pm 0.06$\\
    $L$ (L$_{\odot}$) & $0.08 \pm 0.01$\\
    $T_{\rm eff}$ (K) & $4090 \pm 50$\\[0pt]
    [Fe/H] (dex)     & $0.16 \pm 0.16$\\
    \noalign{\smallskip}
    \hline
    \noalign{\smallskip}
    Orbital Solution \\
    \hline
    \noalign{\smallskip}
    $K$ (m\,s$^{-1}$)                         &  $5.14^{+0.40}_{-0.48}$     \\[1ex]
    $P$ (d)                                   &  $86.54^{+0.07}_{-0.06}$    \\[1ex]
    $e$                                       &  $0.01^{+0.12}_{-0.01}$     \\[1ex]
    $\varpi$ (deg)                            &  $93^{+111}_{-122}$    \\[1ex]
    $M$ (deg)                                 &  $59^{+112}_{-123}$    \\[1ex]
    $t_{\rm trans}$ (MJD)                      &  $2451665.1^{+4.9}_{-4.1}$ \\[1ex]      
    $\gamma_{\rm HIRES}$~(m\,s$^{-1}$)        &  $-1.90^{+0.76}_{-0.80}$    \\[1ex] 
    $\gamma_{\rm CARM.}$~(m\,s$^{-1}$)        &  $0.78^{+0.36}_{-0.32}$     \\ [1ex] 
    $\sigma_{\rm jitter, HIRES}$ (m\,s$^{-1}$)    & $3.7^{+1.0}_{-0.2}$  \\[1ex]
    $\sigma_{\rm jitter, CARMENES}$ (m\,s$^{-1}$) & $2.9^{+0.4}_{-0.17}$  \\[1ex]
    $a$ (au)                                  &  $0.3193^{+0.0002}_{-0.0002}$ \\[1ex]
    $m_{\rm P} \sin{i}$ ($M_{\oplus}$)        &  $24.7^{+1.8}_{-2.4}$      \\[1ex]
    \hline
  \end{tabular}
\end{table}

In Fig.\,\ref{fig:Phase_RV} we show the phased RVs together with the
best fit that includes additional RV jitter.
The total jitter, $(\sigma_{\rm int}^2 + \sigma_{\rm
  jitter}^2)^{1/2}$, is on the order of 3--4\,m\,s$^{-1}$ for both
instruments. The jitter term determined for the HIRES data is
consistent with the term reported by \cite{2010ApJ...725..875I} for
the HIRES data of HD\,147379. We also tried a model fit considering no
RV jitter. All resulting parameters are well within the 1$\sigma$
error bars of those listed in Table\,\ref{table:1} except for the
eccentricity and the argument of periastron. The model including
jitter yields an orbital solution with low eccentricity ($e < 0.13$),
while the model with no jitter results in a rather high eccentricity
($e = 0.29$). We favor the model that considers RV jitter under the
assumption that there is only one detectable planet orbiting the star
and that all the RV noise is a combination of stellar jitter and
instrumental noise systematics.

\begin{figure}
  \includegraphics[width=\hsize, bb = 30 75 625 255, clip=]{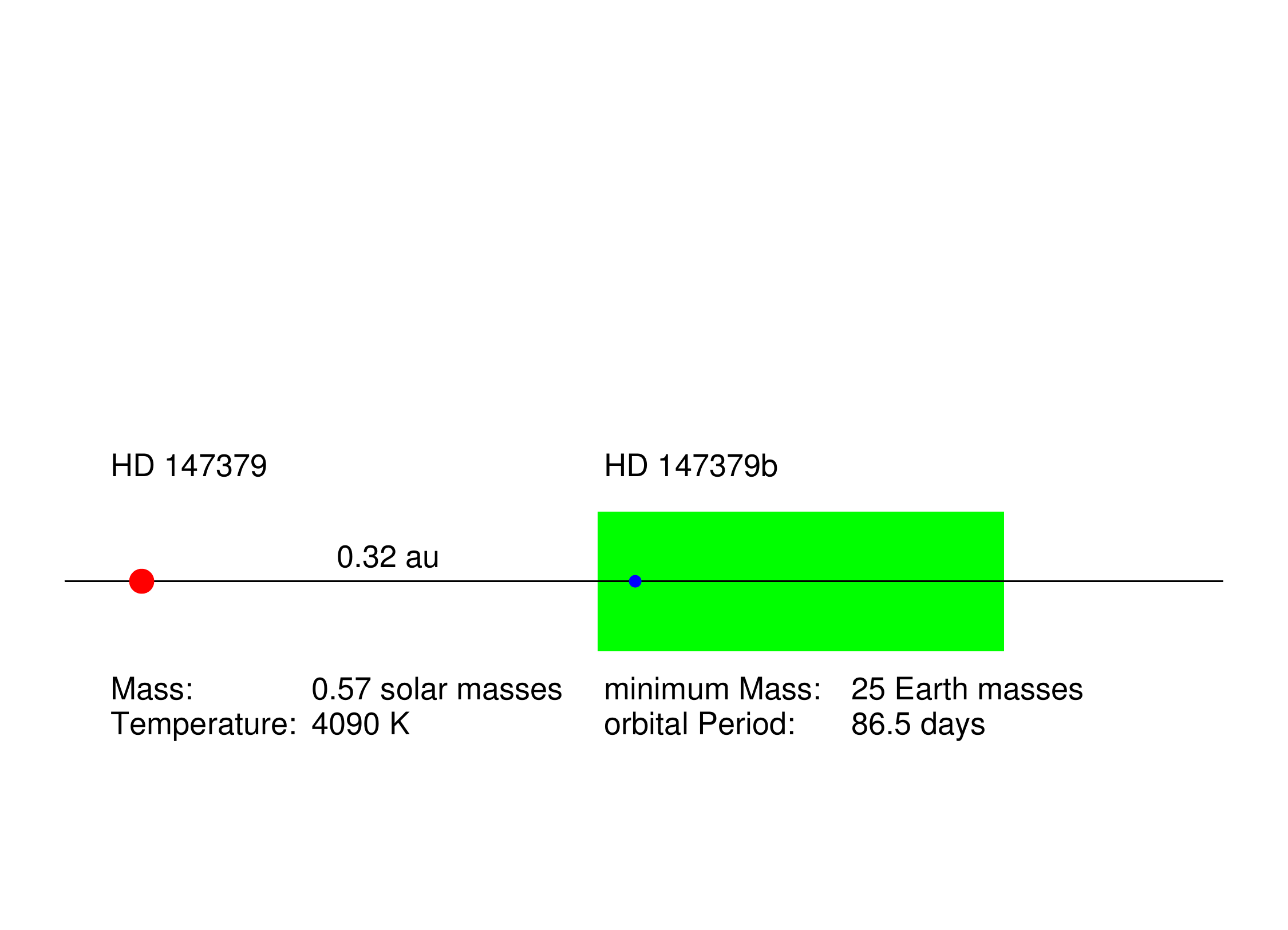}
  \hspace{-.15\hsize} 
  \raisebox{.11\hsize}{\includegraphics[width=.1\hsize, bb = 35 -200 555 705, clip=]{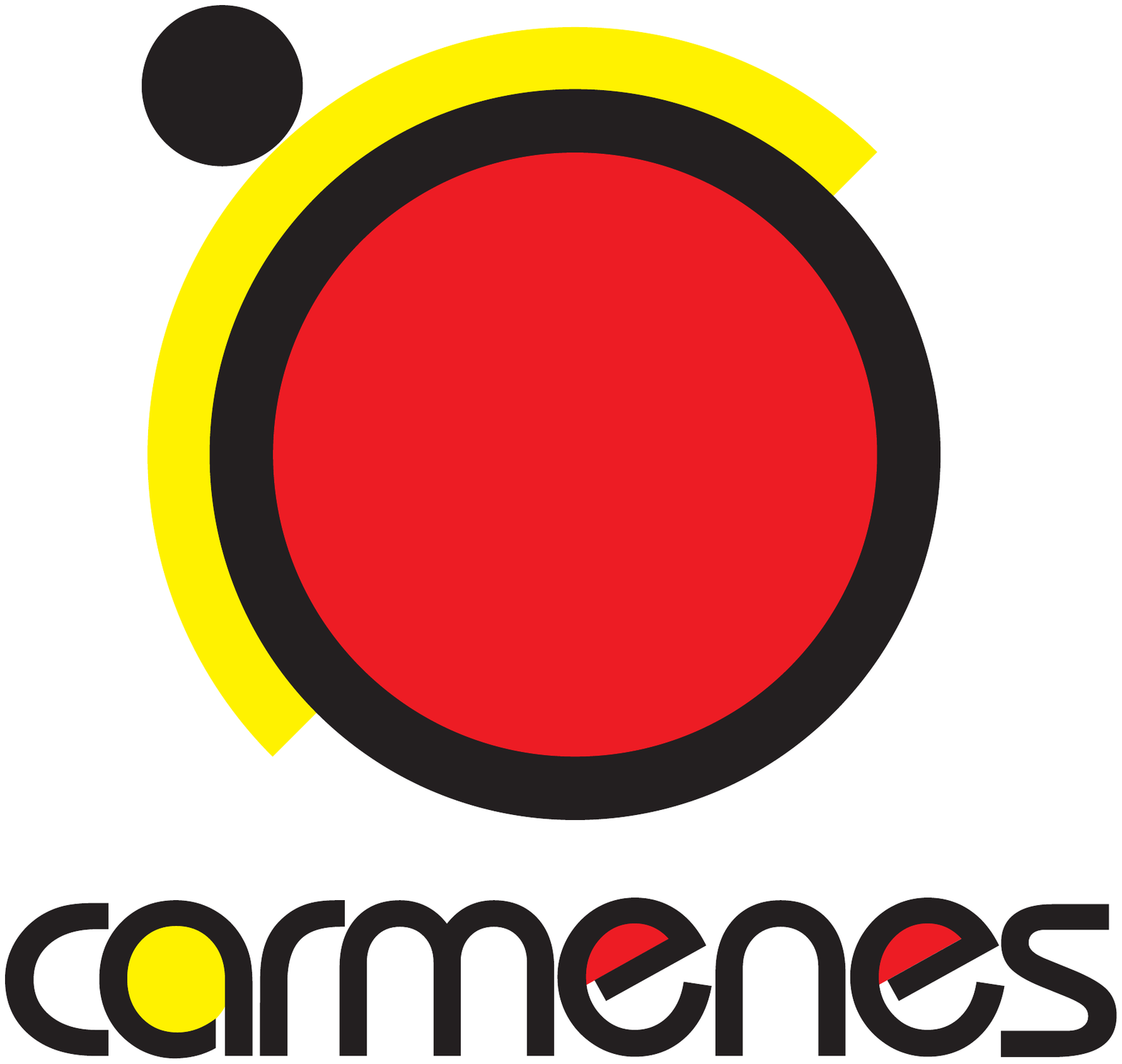}}
  \caption{Visualization of HD\,147379 (red circle) and its planet (blue
    circle).  The green area indicates the conservative habitable-zone limits.}
  \label{fig:Orbit}
\end{figure}

The best-fit orbital solution yields a planetary companion of mass $m_{\rm P}
\sin{i} \approx 25$\,M$_\oplus$ with an orbital semi-major axis $a = 0.32$\,au
and low eccentricity, which locates the planet inside the liquid-water
temperate zone around HD\,147379 \citep{2013ApJ...765..131K,
  2014ApJ...787L..29K}. According to our MCMC posterior distribution, however,
the planetary eccentricity is poorly constrained, and within 2$\sigma,$ we find
that $e < 0.25$. We note that such a high eccentricity for HD\,147379\,b
would cause the planet to approach the star at closer than the habitable-zone
limit near the periastron orbital phase.  The best-fit orbit of HD\,147379\,b
is depicted in Fig.\,\ref{fig:Orbit} together with the limits of the
conservative habitable
zone.\footnote{http://depts.washington.edu/naivpl/sites/default/files/hz.shtml}

After subtracting the orbital motion caused by HD\,147379\,b from the
observed RVs, we calculated the residual periodogram for the CARMENES,
HIRES/Keck, and combined data sets (Fig.\,\ref{fig:Periodogram_res}). The
resulting periodograms show no significant additional periodicities. Our
analysis indicates that any possible further planet in the system should have
an RV amplitude significantly below $\sim$2\,m\,s$^{-1}$.

\section{Conclusions}
\label{sect:conclusions}

Radial velocity observations of HD\,147379 reveal a
25-M$_{\oplus}$ planet in the temperate zone around this early-M
star. HD\,147379 is the first star discovered to host a planet by the
CARMENES search for exoplanets around M dwarfs. The existence of the
planet and its orbital parameters are supported by RV observations
from HIRES/Keck.

With a mass between those of Saturn and Neptune ($m_{\rm
  HD\,147379\,b} \approx$\,0.3\,M$_{\rm Saturn}\approx$\,1.5\,M$_{\rm
  Neptune}$), HD\,147379\,b occupies a mass range that is relatively
poorly populated, especially around stars that are significantly less massive
than the Sun. This planet is located in mass between the super-Earths
that grow large enough to open a gap in the disk for type II migration
but cannot continue accreting, and the ``main clump'' planets that can
trigger runaway gas accretion and rapidly grow larger
\citep{2009A&A...501.1139M}. HD\,147379\,b is similar to the known
planets GJ\,436\,b \citep{2004ApJ...617..580B}, GJ\,3293\,b
\citep{2015A&A...575A.119A}, GJ\,229\,b, and GJ\,433\,c
\citep{2014MNRAS.441.1545T}, but is located inside the temperate zone
of its host star.

The astrometric motion of HD\,147379 is relatively large, and the star
is a good candidate for determining orbital motion with \emph{Gaia} in
a low-mass star. When we assume that the orbit is circular, the lower limit for the
orbital motion semi-amplitude of HD\,147379 that is caused by HD\,147379\,b
and seen from Earth is 4.1\,$\mu$as. If the system is seen under
inclination angles lower than $i = 90$\,deg, the mass of the planet is
higher, and so is the astrometric orbit. The expected performance of
\emph{Gaia} astrometry for HD\,147379 is
6.8\,$\mu$as.\footnote{https://www.cosmos.esa.int/web/gaia/science-performance}
Thus, astrometric detection of the orbital motion of HD\,147379 caused
by its planet is likely possible with \emph{Gaia}.

HD\,147379\,b would be extremely valuable in terms of characterization
potential if it transits its host star. The expected transit depth is of 5--10
mmag, but it has a low geometric transit probability of only 0.8\%.
Photometric follow-up from the ground is complicated by the long period and
correspondingly extended transit duration and the uncertainty of the
conjunction phase. Nevertheless, it lies only 10\,deg away from the ecliptic
pole, and \emph{TESS} should be able to determine whether transits occur.

The discovery of HD\,147379\,b demonstrates the advantage of programs
designed to find planets in orbits of days to months, which are
particularly critical for exploring the habitability zone of M dwarfs. The discovery also shows that some of these planets have likely been missed in previous
searches.

\begin{acknowledgements}
  We thank an anonymous referee for prompt attention and helpful
  comments. CARMENES is an instrument for the Centro Astron\'omico
  Hispano-Alem\'an de Calar Alto (CAHA, Almer\'{\i}a, Spain).  CARMENES is
  funded by the German Max-Planck-Gesellschaft (MPG), the Spanish Consejo
  Superior de Investigaciones Cient\'{\i}ficas (CSIC), the European Union
  through FEDER/ERF FICTS-2011-02 funds, and the members of the CARMENES
  Consortium (Max-Planck-Institut f\"ur Astronomie, Instituto de
  Astrof\'{\i}sica de Andaluc\'{\i}a, Landessternwarte K\"onigstuhl, Institut
  de Ci\`encies de l'Espai, Insitut f\"ur Astrophysik G\"ottingen, Universidad
  Complutense de Madrid, Th\"uringer Landessternwarte Tautenburg, Instituto de
  Astrof\'{\i}sica de Canarias, Hamburger Sternwarte, Centro de
  Astrobiolog\'{\i}a and Centro Astron\'omico Hispano-Alem\'an), with
  additional contributions by the Spanish Ministry of Economy, the German
  Science Foundation through the Major Research Instrumentation Programme and
  DFG Research Unit FOR2544 ``Blue Planets around Red Stars'', the Klaus
  Tschira Stiftung, the states of Baden-W\"urttemberg and Niedersachsen, and
  by the Junta de Andaluc\'{\i}a.  We acknowledge the following funding
  programs: European Research Council (ERC-279347), Deutsche
  Forschungsgemeinschaft (RE 1664/12-1, RE 2694/4-1), Bundesministerium f\"ur
  Bildung und Forschung (BMBF-05A14MG3, BMBF-05A17MG3), Spanish Ministry of
  Economy and Competitiveness (MINECO, grants AYA2015-68012-C2-2-P,
  AYA2016-79425-C3-1,2,3-P, AYA2015-69350-C3-2-P, AYA2014-54348-C03-01,
  AYA2014-56359-P, AYA2014-54348-C3-2-R, AYA2016-79425-C3-3-P and 2013 Ram\`on
  y Cajal program RYC-2013-14875), Fondo Europeo de Desarrollo Regional
  (FEDER, grant ESP2016-80435-C2-1-R, ESP2015-65712- C5-5-R), Generalitat de
  Catalunya/CERCA programme, Spanish Ministerio de Educaci\'on, Cultura y
  Deporte, programa de Formaci\'on de Profesorado Universitario (grant
  FPU15/01476), Deutsches Zentrum f\"ur Luft- und Raumfahrt (grants 50OW0204
  and 50OO1501), Office of Naval Research Global (award no. N62909-15-1-2011),
  Mexican CONACyT grant CB-2012-183007.
\end{acknowledgements}

\bibliographystyle{aa}
\bibliography{refs}

\begin{appendix}

\section{CARMENES radial velocities}

\longtab[1]{
  \begin{longtable}{rr}
    \caption{\label{tab:RVs}Modified Julian Date and CARMENES radial velocities for HD~147379}\\
    \hline\hline
    MJD & $\varv_{\rm rad}$ (m/s)\\
    \hline
    \endfirsthead
    \caption{continued.}\\
    \hline\hline
    MJD & $\varv_{\rm rad}$ (m/s)\\
    \hline
    \endhead
    \hline
    \endfoot
    2457397.691 & $  1.97 \pm  2.62$ \\
2457398.734 & $  0.73 \pm  2.69$ \\
2457400.753 & $ -4.65 \pm  1.61$ \\
2457401.754 & $ -6.53 \pm  1.53$ \\
2457415.688 & $ -5.07 \pm  1.38$ \\
2457419.712 & $  3.58 \pm  1.28$ \\
2457421.735 & $  0.93 \pm  1.14$ \\
2457422.656 & $ -1.97 \pm  1.62$ \\
2457426.748 & $ -2.42 \pm  1.30$ \\
2457427.606 & $ -2.60 \pm  1.52$ \\
2457430.636 & $  3.66 \pm  2.39$ \\
2457436.638 & $  1.01 \pm  2.07$ \\
2457440.635 & $  6.01 \pm  1.49$ \\
2457441.761 & $  8.71 \pm  1.53$ \\
2457444.691 & $  2.31 \pm  1.57$ \\
2457466.729 & $ -7.41 \pm  1.61$ \\
2457472.672 & $ -2.78 \pm  1.31$ \\
2457490.622 & $ -8.01 \pm  1.24$ \\
2457495.612 & $ -1.86 \pm  1.93$ \\
2457505.552 & $  3.00 \pm  1.45$ \\
2457509.598 & $ -0.66 \pm  1.26$ \\
2457529.506 & $  4.01 \pm  1.56$ \\
2457537.597 & $ -2.50 \pm  2.21$ \\
2457537.609 & $ -2.66 \pm  2.35$ \\
2457539.486 & $ -4.16 \pm  1.67$ \\
2457542.512 & $ -0.99 \pm  1.84$ \\
2457550.436 & $ -3.26 \pm  2.77$ \\
2457550.446 & $ -5.28 \pm  2.30$ \\
2457553.441 & $ -2.28 \pm  2.06$ \\
2457555.545 & $ -1.53 \pm  1.67$ \\
2457596.397 & $  7.82 \pm  3.17$ \\
2457627.374 & $ 11.32 \pm  4.64$ \\
2457646.299 & $ -4.88 \pm  1.91$ \\
2457647.334 & $ -2.86 \pm  2.48$ \\
2457760.750 & $ -7.90 \pm  1.69$ \\
2457761.734 & $ -2.01 \pm  1.74$ \\
2457768.751 & $ -3.68 \pm  2.05$ \\
2457779.650 & $  4.69 \pm  1.91$ \\
2457791.624 & $  7.73 \pm  2.30$ \\
2457792.653 & $  2.99 \pm  2.76$ \\
2457793.614 & $  5.17 \pm  1.95$ \\
2457798.605 & $  0.46 \pm  1.45$ \\
2457799.659 & $  2.48 \pm  2.12$ \\
2457806.595 & $  6.18 \pm  2.48$ \\
2457815.733 & $ -3.49 \pm  1.68$ \\
2457817.621 & $ -0.44 \pm  2.42$ \\
2457819.635 & $ -2.29 \pm  1.87$ \\
2457824.500 & $ -5.65 \pm  1.74$ \\
2457828.575 & $ -4.29 \pm  1.49$ \\
2457829.561 & $ -8.51 \pm  1.77$ \\
2457830.634 & $ -6.02 \pm  1.54$ \\
2457833.562 & $ -1.33 \pm  1.82$ \\
2457834.691 & $ -0.15 \pm  1.79$ \\
2457852.703 & $ -3.23 \pm  1.28$ \\
2457855.608 & $ -0.50 \pm  1.30$ \\
2457857.475 & $  0.54 \pm  1.34$ \\
2457858.579 & $  0.87 \pm  1.92$ \\
2457859.513 & $  2.38 \pm  2.27$ \\
2457860.486 & $ -0.07 \pm  1.23$ \\
2457863.518 & $  2.86 \pm  2.20$ \\
2457864.588 & $  4.87 \pm  1.48$ \\
2457866.595 & $  5.83 \pm  1.62$ \\
2457875.474 & $  3.05 \pm  1.55$ \\
2457877.458 & $  5.35 \pm  1.54$ \\
2457878.651 & $  3.34 \pm  3.48$ \\
2457879.447 & $  8.15 \pm  4.28$ \\
2457882.548 & $  3.35 \pm  1.39$ \\
2457883.466 & $  2.63 \pm  1.83$ \\
2457887.560 & $  4.19 \pm  1.50$ \\
2457888.486 & $  4.15 \pm  1.47$ \\
2457889.462 & $  0.70 \pm  1.80$ \\
2457890.516 & $ -1.76 \pm  1.66$ \\
2457894.539 & $ -5.00 \pm  1.59$ \\
2457898.555 & $ -4.10 \pm  1.74$ \\
2457901.431 & $ -2.69 \pm  1.97$ \\
2457905.484 & $ -3.92 \pm  1.62$ \\
2457907.447 & $ -7.39 \pm  1.64$ \\
2457909.487 & $ -7.25 \pm  2.07$ \\
2457910.522 & $-12.36 \pm  2.15$ \\
2457912.536 & $ -9.22 \pm  1.94$ \\
2457914.540 & $ -8.27 \pm  1.79$ \\
2457915.479 & $ -8.64 \pm  1.83$ \\
2457916.443 & $ -9.06 \pm  2.15$ \\
2457919.364 & $ -9.55 \pm  2.59$ \\
2457920.560 & $ -8.69 \pm  2.58$ \\
2457921.435 & $ -4.30 \pm  1.84$ \\
2457922.478 & $ -3.28 \pm  1.68$ \\
2457928.374 & $  0.24 \pm  1.73$ \\
2457934.424 & $ -2.08 \pm  2.46$ \\
2457935.443 & $  0.86 \pm  1.59$ \\
2457937.365 & $  1.21 \pm  1.58$ \\
2457942.496 & $  2.09 \pm  1.82$ \\
2457943.494 & $  0.55 \pm  1.91$ \\
2457945.397 & $  4.58 \pm  1.49$ \\
2457946.399 & $  5.30 \pm  1.72$ \\
2457947.432 & $  5.86 \pm  1.72$ \\
2457949.540 & $  2.22 \pm  1.49$ \\
2457950.548 & $ -1.26 \pm  1.54$ \\
2457951.431 & $ -0.58 \pm  1.45$ \\
2457954.352 & $  4.22 \pm  1.87$ \\
2457956.438 & $  9.34 \pm  1.71$ \\
2457960.373 & $  3.92 \pm  1.31$ \\
2457961.368 & $  2.38 \pm  1.43$ \\
2457962.425 & $  6.71 \pm  1.37$ \\
2457963.373 & $  6.90 \pm  1.56$ \\
2457965.352 & $  9.58 \pm  1.52$ \\
2457968.395 & $ 11.65 \pm  1.47$ \\
2457980.374 & $  1.67 \pm  2.79$ \\
2457981.388 & $ -3.54 \pm  1.68$ \\
2457982.513 & $ -1.40 \pm  1.76$ \\
2457985.463 & $  3.68 \pm  1.97$ \\
2457987.489 & $  7.83 \pm  1.79$ \\
2457989.331 & $  8.33 \pm  2.54$ \\
2457999.464 & $  2.12 \pm  2.69$ \\

  \end{longtable}
}

\section{Residuals}

\begin{figure*}
  \centering
  \includegraphics[width=\hsize, bb = 0 90 640 465, clip=]{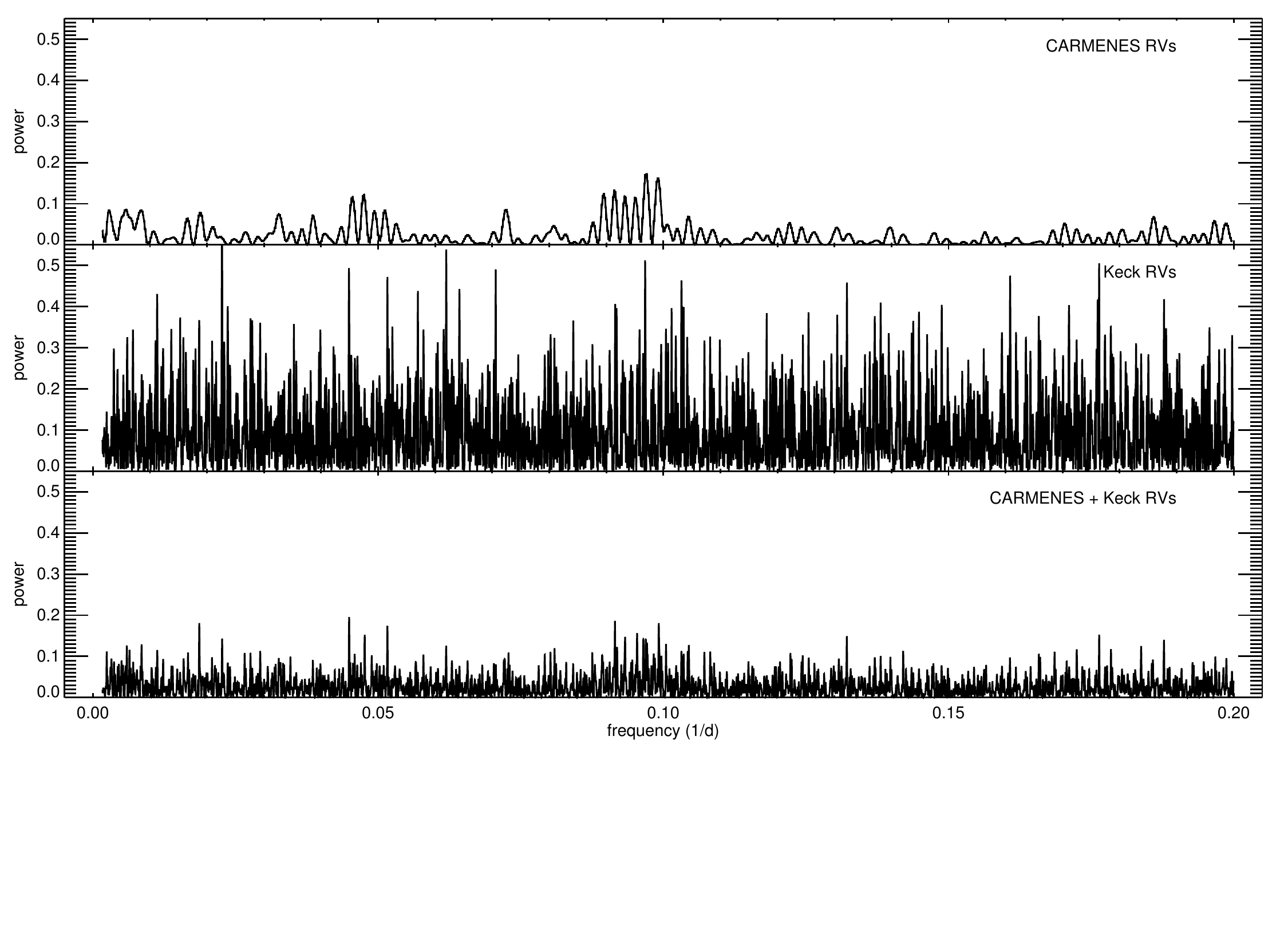}
  \caption{Periodogram of residuals after removing the 86.5\,d planet signal.}
  \label{fig:Periodogram_res}
\end{figure*}

\end{appendix}

\end{document}